\documentclass[12pt]{iopart}
\usepackage{graphicx}
\usepackage{amssymb}
\usepackage{mathrsfs}
\usepackage{amsfonts}
\usepackage{dsfont}
\usepackage{dcolumn}
\usepackage{bm}
\usepackage{booktabs}
\usepackage{textcomp}
\usepackage{color}
\usepackage[usenames,dvipsnames]{xcolor}
\usepackage{bbold}
\usepackage{setstack}
\usepackage{paralist}

\newcommand{\hr}{\hat{\rho}}
\newcommand{\hU}{{\hat{U}}}

\providecommand{\abs}[1]{\left|#1\right|}

\providecommand{\ket}[1]{|#1\rangle}

\providecommand{\abs}[1]{\left\lvert#1\right\rvert}

\providecommand{\ket}[1]{|#1\rangle}

\begin{document}

\title{A retrodiction paradox in quantum and classical optics}

\author{Andrea Aiello$^{1,2}$ and J. P. Woerdman$^{3}$}

\address{$^1$ Max Planck Institute for the Science of Light, G$\ddot{u}$nther-Scharowsky-Strasse 1/Bau24, 91058
Erlangen,
Germany}
\address{$^2$Institute for Optics, Information and Photonics, University of Erlangen-Nuernberg, Staudtstrasse 7/B2,
91058 Erlangen, Germany}
\address{$^3$ Huygens-Kamerlingh Onnes Laboratory, Leiden University,
P.O. Box 9504, 2300 RA Leiden, The Netherlands}

\begin{abstract}
Quantum mechanics represents one of the greatest triumphs of human intellect and, undoubtedly, is  the most successful physical theory we have to date. However,  since its  foundation  about a century  ago, it has been uninterruptedly the center of harsh debates ignited by the counterintuitive  character of some of its predictions. The subject of one of these heated discussions is the so-called ``retrodiction paradox", namely a deceptive inconsistency of quantum mechanics which is often associated  with  the ``measurement  paradox"  and the ``collapse of the wave function";  it comes from the apparent time-asymmetry between state preparation and measurement.  Actually, in the literature one finds several versions of the retrodiction paradox; however, a  particularly insightful one was presented by Sir Roger Penrose in his seminal book \emph{The Road to Reality}.
 Here, we address the question to what degree  Penrose's retrodiction paradox  occurs in the classical  and quantum domain.    We achieve a twofold result. First, we show that Penrose's  paradox manifests itself in some form also in classical optics.  Second, we demonstrate that when information is correctly extracted from the  measurements and  the quantum-mechanical formalism is properly applied, Penrose's retrodiction paradox does not manifest itself in quantum optics.
\end{abstract}




\section{Introduction}\label{Introduction}

\begin{quote}
``\emph{It seems to me that there are deep philosophical lessons to be learned in the way in which the practicing theoretical physicist thinks about the foundations of the subject} [...] \emph{So, the important thing then is to display the general world view, the world picture that the theoretical physicists has.}'' (Julian Schwinger \cite{Schwinger}, Prologue, page 1).
\end{quote}

According to the Oxford Dictionary, \emph{retrodiction} is ``the explanation or interpretation of past actions or events inferred from the laws that are assumed to have governed them'' \cite{Oxford}. Oppositely, \emph{prediction} denotes a statement about future events based on the knowledge of past occurrences. Predictions and retrodictions are commonplace in everyday life. We make predictions when we place a bet on a horse; when we arrange the time for a date; when we
 forecast a heavy rainfall for the weekend;  and when we invest in the stock market.
 Retrodictions are made when a sommelier infers the production year of a wine from tasting; when a detective closes a case on the ground of the gathered evidences; when a radio operator reconstructs a message from a Morse code; and when we guess the production cost of an item from its quality.

 Alas, things do not go so smoothly when dealing with predictions and retrodictions in the realm of  physics.
 The problems are rooted in the profoundly different descriptions of nature provided by classical and quantum physics.
The natural  world as perceived by our senses is ruled by the laws of classical physics established  by Galileo, Newton, Maxwell,  and few others over the past  centuries. However, the microscopic world of atoms, {photons} and elementary particles obeys the strange laws of quantum mechanics formulated about a century ago by Planck, Einstein, Schr\"{o}dinger, Heisenberg et alii. Thus, classical and quantum mechanics offer two different, and often irreconcilable, representations of the physical world. A consequence thereof, is the rise of  deceptive inconsistencies in the quantum mechanical formalism when viewed through the lens of classical mechanics, see e.g.  \cite{Peres,Griffiths,Vaidman}. One of such ``flaws'' is the so-called retrodiction paradox due to the apparent time-asymmetry in quantum state reduction.

\begin{figure}[!h]
%
\centerline{\includegraphics[scale=3,clip=false,width=.8\columnwidth,trim = 0 0 0 0]{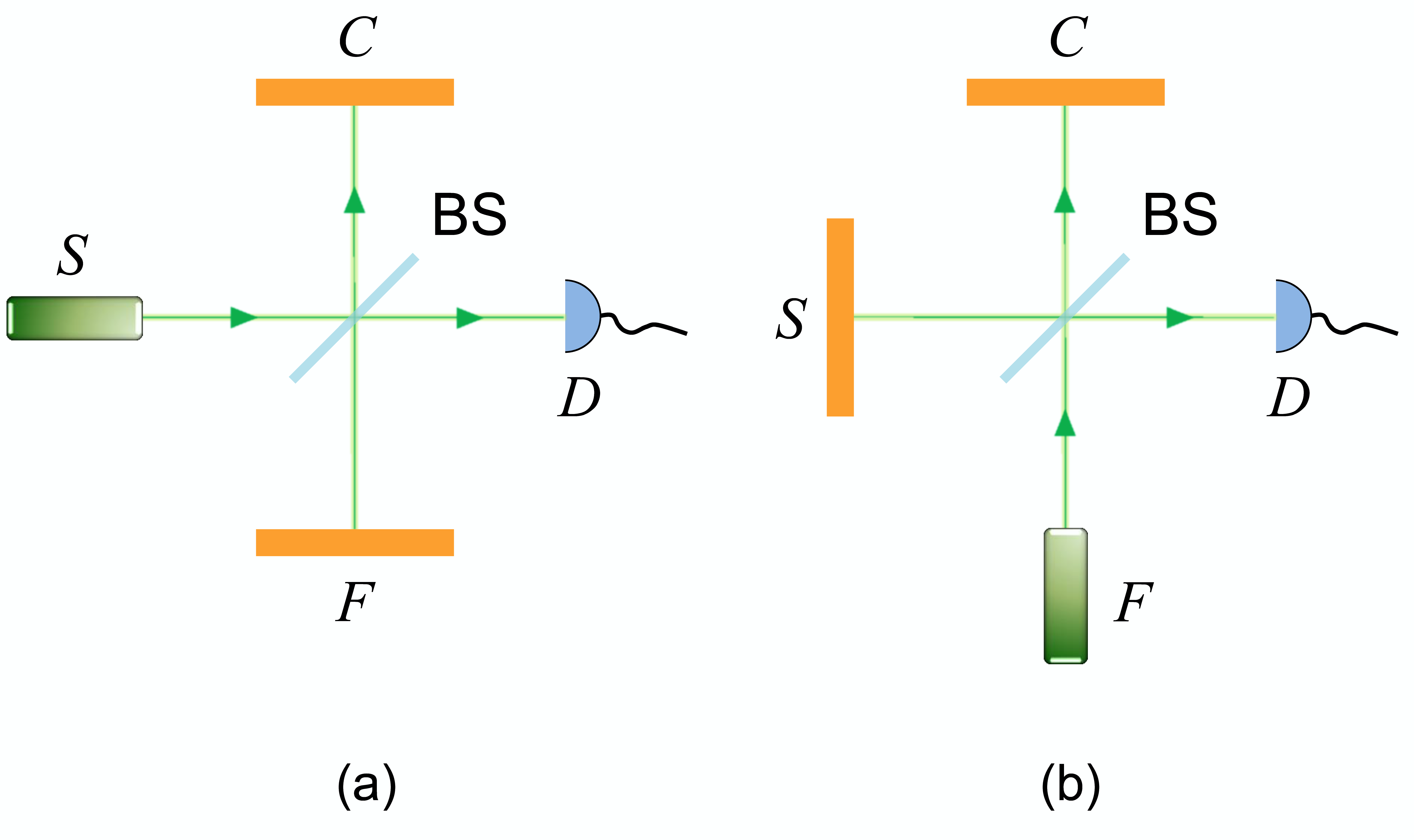}}
\caption{\label{fig1}
Penrose's version of the retrodiction paradox in quantum optics. (a)  A random photon source $S$ emits one photon per time in the direction of a photodetector $D$. Midway between $S$ and $D$ the ${50\!:\!50}$ beam-splitter $\mathsf{BS}$ either transmits or reflects the  photon with equal probability $1/2$. If the photon is transmitted, it drives the detector $D$ to click; if reflected, the photon is eventually absorbed by the ceiling $C$. Quantum mechanics correctly  \emph{predicts} the same  $50\%$ probability for both events.  (b) Hypothetical situation where the source $S$ and the floor $F$ had exchanged their roles. The  source $F$ emits one photon aimed at the beam splitter $\mathsf{BS}$.
If reflected by  $\mathsf{BS}$,  it triggers the detector $D$; if transmitted, the photon is then absorbed by the ceiling $C$. As before, quantum mechanics  predicts   $50\%$ probability for either occurrences.
Cases (a) and (b) illustrate two \emph{distinct} physical systems. Nonetheless, in both instances the photon ends up in  either the detector or in the ceiling. Therefore, whenever $D$ counts one photon we could infer that the source was located either as in (a) with $50\%$ probability or as in (b) with again $50\%$ probability.
However, such straightforward inference seems to contradict factual events because by definition only case (a) actually occur in reality, being case (b) purely hypothetical: this yields to the quantum retrodiction paradox as explained in the main text.
}
\end{figure}

A facet of this interesting problem has been vividly illustrated {in the context of quantum optics by Sir Roger Penrose in his bestseller book \emph{The Road to Reality} \cite{Penrose}, as follows}. Consider a light source $S$ emitting one photon towards a photodetector $D$, as shown in figure \ref{fig1} (a).
Midway between $S$ and $D$ the  ${50\!:\!50}$ beam splitter $\mathsf{BS}$ can either transmit or reflect the incoming photon with equal probability $1/2$. In the former case the photon reaches the detector $D$ and is recorded. In the latter case the photon hits the ceiling  $C$ and is absorbed. Given that the source had emitted a photon at time $t=0$, quantum mechanics \emph{predicts} that at a sufficiently later time $t=T>0$ there is $50\%$  probability of a detection event at $D$ and $50\%$ probability of a photon being absorbed by $C$. So far so good. However, according to Penrose, a ``retrodiction paradox'' arises when considering the time-reversed process and posing the question:
\begin{quote}
\emph{Given that a photon has been detected by  $D$ at some time $t=T$, what is the probability of an  early  emission event by  $S$ at a sufficiently former time $t=0$?}
 \end{quote}
 The only reasonable answer  seems to be: ``$100\%$'', because $S$ is the only available source. If this probability were not $100\%$ we would implicitly admit that the detected photon could have been emitted by the floor $F$ and ``this, of course, is an absurdity'', in Penrose's words.
Nonetheless,  according to Penrose, quantum mechanics precisely  \emph{retrodicts} that there is $50\%$  probability that the detected photon was emitted by the source $S$ and $50\%$ probability for emission by the floor $F$. Why?

\begin{figure}[!ht]
%
\centerline{\includegraphics[scale=3,clip=false,width=.4\columnwidth,trim = 0 0 0 0]{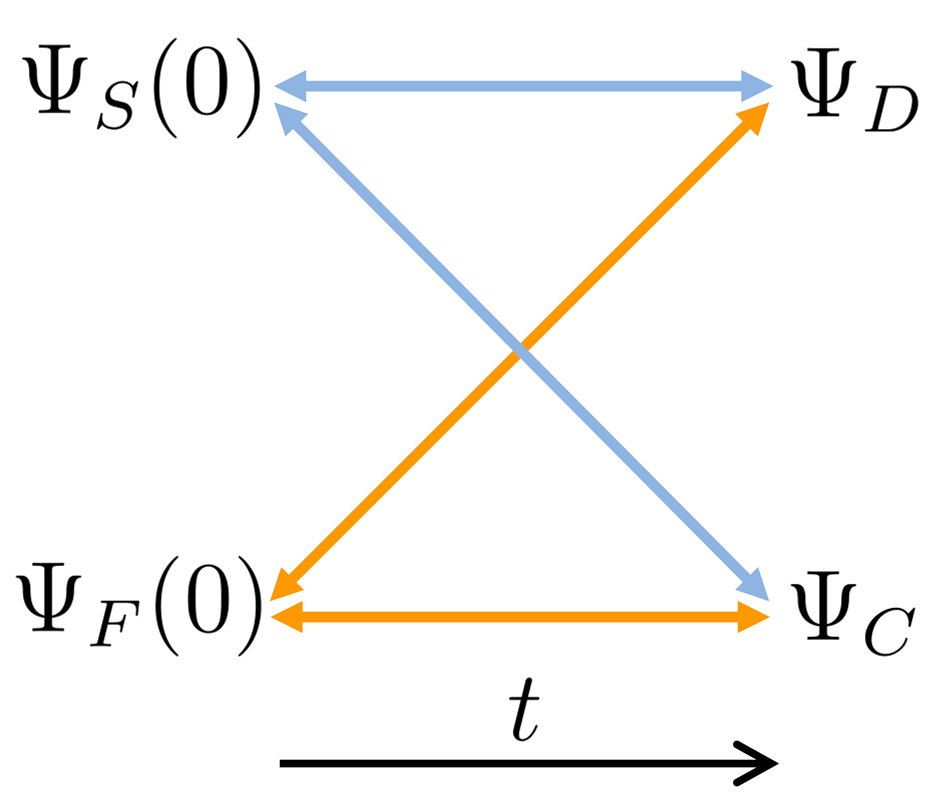}}
\caption{\label{fig2}
Schematic illustration of the  time-evolution of the state vectors $\Psi_S(0)$  and $\Psi_F(0)$. As time $t$ goes on, the state $\Psi_S(0)$ ends up in a equal-probability superposition of $\Psi_D$  and  $\Psi_C$. However, given the final state $\Psi_D$, it can be traced back to either $\Psi_S(0)$ (light-blue arrow)  or $\Psi_F(0)$ (light-orange arrow) with the same probability of $50\%$. The same reasoning applies to $\Psi_F(0)$.}
\end{figure}

If we content ourselves with a \emph{formal} answer, this can be easily given as follows.
To begin with, let us recall that
the first postulate of quantum mechanics is that under certain conditions a physical system can be represented by a \emph{state vector}\footnote{Throughout this work we adhere to Peres' view about the \emph{physical} meaning of the state vector \cite{Peres3}: ``A state vector is not a property of a physical system (nor of an ensemble of systems). It does not evolves continuously between measurements, nor suddenly ``collapse'' into a new state vector whenever a measurement is performed. Rather, a state vector represents a \emph{procedure} for preparing or testing one or more physical systems.'' See also \cite{Newton}.}
in an abstract linear space  known as \emph{Hilbert space} \cite{Weinberg}.
 Thus, let $\Psi_S(0)$ be the state vector associated to the  photon emitted by $S$ at time $t=0$\footnote{Following  \cite{Weinberg}, in this paper we use capital Greek letters $\Psi, \Phi, \Omega,\ldots$ to denote quantum state vectors.}.
 Applying the standard rules of quantum mechanics it is not difficult to show  that at  later time $t=T>0$  the state vector of the photon is
\begin{eqnarray}\label{eq10}
\Psi_S(T) = (\Psi_D + i \, \Psi_C)/\sqrt{2},
\end{eqnarray}
 where $\Psi_D$ represents the photon aimed at $D$ and  $\Psi_C$ describes the photon directed at $C$. Both states have the same probability $1/2$ of occurrence. Now, consider the hypothetical situation depicted in figure \ref{fig1} (b) where the source is located on the floor $F$. Let  $\Psi_F(0)$ be the state vector representing the photon emitted by this source at $t=0$.
 Proceeding as before, at  $t=T>0$ the state of the photon will be
\begin{eqnarray}\label{eq20}
\Psi_F(T) = (i \, \Psi_D +  \Psi_C)/\sqrt{2},
\end{eqnarray}
  where $\Psi_D$  and $\Psi_C$ are given as above. A pictorial representation of $\Psi_S$  and $\Psi_F$ and  their connections with the vectors $\Psi_D$  and $\Psi_C$ is given in figure \ref{fig2}.
\\
As shown in this illustration, given at $t=0$ the \emph{unique} initial vector $\Psi_S(0)$, there are \emph{two} different state vectors $\Psi_D$  and $\Psi_C$  associated to it at  $t=T$ (light-blue forward arrows)\footnote{This is peculiar to quantum mechanics: Given a physical system initially prepared in a specific and unique state, as time goes on it may evolve in a superposition of \emph{several} distinct states, each occurring with an assigned  probability \cite{PeresBook,Jordan}.}.
 However, when going backwards in time, to each state vector $\Psi_D$ and $\Psi_C$ are associated \emph{both}  input states  $\Psi_S(0)$  and $\Psi_F(0)$ (light-blue and light-orange backward arrows). Of course, $\Psi_S(0)$  and $\Psi_F(0)$ describe two very distinct physical situations:  $\Psi_S(0)$  represents the state of the photon emitted by the source at $S$ (figure \ref{fig1} (a)), while $\Psi_F(0)$   describes the state of the photon emitted by the source at $F$ (figure \ref{fig1} (b)). Nonetheless, both vector states end up at $t= T$ into a $50 \slash 50$ superpositions of  $\Psi_D$  and $\Psi_C$ (with different phases).
This simply means that irrespective of the location of the source, the emitted photon has $50\%$  probability of exciting the detector $D$ and $50\%$ probability of  being absorbed by $C$. Therefore,  { given \emph{the sole information} that a photon has been detected by $D$, it seems that} quantum mechanics cannot retrodict correctly the position of the source, namely cannot discriminate between $\Psi_S(0)$  and $\Psi_F(0)$. This is the formal reason for the peculiar quantum mechanical retrodiction  for the present problem.

In this work we aim at giving a substantial, as opposed to formal, answer to the question posed by Penrose's quantum retrodiction paradox. Surprisingly enough, we find that the substantial answer  strongly differs from the formal one, thus revealing that, de facto, \emph{there is no retrodiction paradox} in the form presented by Penrose. We begin our study by showing in the next section that, according to d'Espagnat \cite{Espagnat}, a retrodiction paradox may originate from an incorrect handling of the information obtained from  measurement processes.
Then,
  in the third section we  show that some of the alleged inconsistencies in  Penrose's quantum optics retrodiction problem already deceptively appear in the corresponding classical optics retrodiction problem.  Finally, by means of a concrete and fully developed example, in the penultimate section we  demonstrate that, contrarily to Penrose's claim, quantum mechanics does indeed correctly retrodict the position of the light source, provided that \emph{enough information} about the system is correctly extracted from the measurements. In the last section we draw our conclusions

\section{Measurement and information}\label{Unitary}

\begin{quote}
``\emph{The problem of retrodiction is associated with basic questions, but for this very reason it is, unfortunately, quite difficult, even in classical physics. In quantum physics it is even more subtle.}'' (Bernard d'Espagnat \cite{Espagnat}, chapter 13, page 149).
\end{quote}

In the previous section we have (erroneously) learned that the quantum retrodiction paradox is rooted in the assignment of the same $50\%$ probability to both the time-ordered process where the photon goes from the source $S$ to the detector $D$ and the time-reversed one with the photon going backward from the detector to the source. How are these equal probabilities actually calculated and why did we obtain a wrong result?  In quantum mechanics, the probability that a system described by the state vector $\Psi$ can be found in the state $\Phi$ is simply given by the modulus square of the so-called \emph{probability amplitude} $\left( \Phi,\Psi \right)$:
\begin{eqnarray}\label{symmetry}
\abs{\left( \Phi,\Psi \right)}^2 = \abs{\left( \Psi,\Phi \right)}^2,
\end{eqnarray}
where the complex amplitude $\left( \Phi,\Psi \right)= \left( \Psi,\Phi \right)^*$ (the asterisk symbol ``$\,{}^*\,$'' denotes complex conjugation) is defined as the scalar product between the vectors $\Psi$ and $\Phi$ which are supposed to  have unit norm, namely $\left( \Psi,\Psi \right)=\left( \Phi,\Phi \right)=1$ \cite{Weinberg}. Moreover, state vectors  $\Phi$ and  $\Omega$ associated to mutually exclusive events are \emph{orthogonal}, that is $\left( \Phi,\Omega \right)=0$. For example, the photon in figure \ref{fig1} (a) is either aiming at $D$ or directed to the ceiling $C$ and, consequently, $\left( \Psi_D,\Psi_C \right)=0$. Therefore, according to these rules and using \eref{eq10}, the probability  that the photon emitted at $t=0$ from the source $S$ will be detected by detector $D$ at $t =T$ is straightforwardly calculated as
\begin{eqnarray}
\abs{\left( \Psi_D,\Psi_S(T) \right)}^2 & = \; & \frac{1}{2}\abs{\left( \Psi_D,\Psi_D \right) +i \left( \Psi_D,\Psi_C \right)}^2 \nonumber \\
& = \; & \frac{1}{2}\abs{1 +i \times 0}^2 =50\%.
\end{eqnarray}
Analogously,  given that a photon has been detected by $D$ at time $T$, the probability that it was emitted by $S$ is by definition
\begin{eqnarray}\label{flaw}
\abs{\left( \Psi_S(T), \Psi_D \right)}^2 = \abs{\left( \Psi_D,\Psi_S(T) \right)}^2 =50\%,
\end{eqnarray}
where the symmetry-property \eref{symmetry} has been used. Therefore, quantum-mechanical calculations seem to confirm the existence of Penrose's paradox. Then, where is the flaw? Well, the problem arises from equation \eref{flaw}, when we assign the state $\Psi_D$ to the photon as consequence of the fact that it has been detected by $D$. As remarked by d'Espagnat (sec. 13.2, p. 155 of \cite{Espagnat}), from the sole information of a detection event at $D$, we cannot infer that the photon was in the state $\Psi_D$ already before detection; this is simply an incorrect use of such information. As a matter of fact, after the beam-splitter $\mathsf{BS}$ the photon becomes a nonlocal object, namely it is not either reflected \emph{or} transmitted by $\mathsf{BS}$; rather it is both reflected \emph{and} transmitted. If we disregard that part of information contained in the state vector $\Psi_C$ describing the reflected photon, we cannot hope to make a correct retrodiction. The same is trivially true in classical mechanics: If many people placed in different positions fire simultaneously a bullet into a target (think of a firing squad), from the sole position of a bullet stuck in the target, we cannot infer who had fired it. However, if we had more information, such as the direction of the flying bullet before being stuck into the target, we could make a correct retrodiction.

In section \ref{Theory} we shall show how to calculate correctly all the quantum-mechanical probabilities relative to our problem and it will become evident that  {Penrose's} retrodiction paradox  disappear.

\section{Classical and quantum representations of physical phenomena}\label{Classical}

\begin{quote}
``\emph{Quantum phenomena do not occur in a Hilbert space. They occur in a laboratory.}'' (Asher Peres \cite{PeresBook}, chapter 12, page 373).
\end{quote}

The aim of this section is twofold. First, for didactic reasons, we intend to introduce the reader to some of the subtleties of the classical and the quantum representations of physical events. Second, we want to show that
 `` {Penrose's} quantum retrodiction paradox'' illustrated in the previous section {in the context of quantum optics}, also becomes {deceptively} manifest at the level of classical optics.

\subsection{Classical or quantum world?}\label{CandQ}

Classical and quantum mechanics offer two different representations of natural world. Nonetheless,  a physical system \emph{per se} should not be regarded as either inherently classical or quantum. Rather, is the way we interact with the system that determines how we represent it.
As a matter of fact, underlying the concepts of classical and quantum physics, there is the notion that a clear distinction between a system obeying the laws of classical physics and one governed by quantum mechanics, can be made. However,  it is not always clear (sometime even meaningless), where to draw the conceptual line that separates the classical  representation of a physical system from the quantum one \cite{Englert}.
Consider, for example, a bottle of volume $V$ containing hydrogen gas at pressure $P$ and temperature $T$. In these conditions, the gas may be regarded as  a \emph{classical} thermodynamic system whose behavior is ruled by an equation of state of the form
\begin{eqnarray}\label{tre210}
f (P,V,T ) =0.
\end{eqnarray}
This equation can be deduced by assuming that  the hydrogen gas obeys the laws of classical physics without actually referring to the atomic constituents of the gas. Moreover, the relation $f (P,V,T ) =0$ can be experimentally verified by actually measuring pressure and temperature of the gas in the bottle with ordinary measuring apparatuses. Imagine now to take the same bottle and to use the gas as a sample for a spectroscopic experiment aimed at determining  the emission spectrum of atomic hydrogen. Of course, the result of such experiment would be the ``discovery'' of a discrete spectrum consisting of a numerable set of isolated spectral lines. Notably, this characteristic of the spectrum of the hydrogen gas  can be predicted only by using the laws of \emph{quantum} physics. Therefore, we arrived at the apparent contradiction of having a physical system, the bottle filled with hydrogen gas, obeying \emph{both} the laws of classical and quantum physics. Where is  the border between classical and quantum physics here?

The lesson to be learned from this simple example is that a unique physical system may admit diverse representations  all equally valid. Thence,  our bottle of hydrogen gas may be regarded either as a classical system if we are concerned with its thermodynamical properties,  or as a quantum system if we are interested to the emission spectrum of atomic hydrogen. Of course, the nature of the system is the same in both cases, it is only the way we look at it (interaction) that requires either a classical or a quantum description \cite{Mermin}.

\subsection{The retrodiction paradox in classical optics}\label{retroclassical}

In this subsection we provide for a classical-optics interpretation of  Penrose's retrodiction paradox,  where the single-photon quantum source $S$ in figure \ref{fig1} (a) is  replaced by a bright classical source as, e.g., a slide projector  or a photographic flash.
According to Penrose, the quantum retrodiction paradox is due to the inherent time-asymmetry in what he calls the ``\textbf{R} procedure'', which is a name for the \emph{state-vector reduction} or \emph{collapse of the wavefunction} (see section 22.1 of \cite{Penrose}). In classical optics, there is not such a thing as the \textbf{R} procedure because
states of light admitting a classical description are represented as smoothly and continuously varying waves. Therefore, we could justifiably  expect that a retrodiction paradox cannot become manifest in the realm of classical optics. Surprisingly enough, here we demonstrate  this is not the case because Penrose's reasoning leading to the quantum paradox can be mostly repeated step by step in the classical case.

To this end, consider again the system illustrated in figure \ref{fig1} (a) and imagine to replace the single-photon  quantum source at $S$ with a still random but bright classical source, say a photographic flash. Then, suppose that a burst of light  is emitted towards the detector $D$. At beam-splitter \textsf{BS}, $50\%$ of the light will be transmitted in the direction of the detector $D$ and $50\%$  will be reflected to the ceiling $C$. Therefore, whenever there is an emission event at $S$ with intensity $I$, there must be  an intensity $I/2$ registered by detector $D$ and an intensity  $I/2$ absorbed by the ceiling $C$. This is what classical optics and experiments tell us. Now, quoting Penrose, ``[...] \emph{let us imagine reading this particular experiment backwards in time.}'' So, assume that the detector $D$ had just recorded a flash of light of intensity $I/2$. Then the relevant question is: where does this light come from? Evidently, it cannot come from the ceiling $C$ because light emitted by the ceiling is either reflected towards the source or transmitted in the direction of the floor $F$. The only remaining possibilities are that  \emph{a})  light of intensity $I$ was emitted by the floor $F$; and \emph{b}) light of intensity $I$ was emitted by the source $S$\footnote{Following Penrose, we are deliberately ignoring the possibility of a \emph{simultaneous} emission by both the floor $F$ and the source $S$. If this were the case, we should have considered two additional instances, accounting for when \emph{c})  $F$ and $S$  emitted together \emph{incoherent}   light of intensity $I/2$; and \emph{d}) $F$ and $S$ emitted simultaneously  \emph{coherent}  light of intensity $I/4$.}. Both of these two possibilities must have  $50\%$ probability of occurrence like in the quantum case because, as  remarked by Penrose,  the ratio of the intensities of the transmitted and reflected light is just an intrinsic property of the beam splitter,  irrespective of the either quantum or classical nature of light. Thus, remarkably, classical optics led us to exactly the same conclusions as  quantum optics!

Of course, in the classical-optics case no one would seriously consider the hypothesis that the flash was emitted by the floor (who pays for the bill?) On the other hand, if we imagine to propagate backwards from the detector a packet of light of intensity $I$, this would certainly split at \textsf{BS} in two packets of intensity $I/2$ each, one directed towards $S$ and the other aimed at $F$. However, this cannot be correct because in the time-forward motion there was not light emitted from $F$. This apparent inconsistency of the classical-optics (and quantum too!) description,  is actually a trivial consequence of our \emph{incomplete} reconstruction of the motion of light. If we had  propagated backwards \emph{also} the light absorbed by the ceiling (which, though, we had not detected), this would have summed \emph{coherently} at \textsf{BS} with the light coming from $D$. As a result of this coherent process, we would have had constructive light interference in the direction of the source, and destructive  interference towards the floor.
This completely solves the classical-optics  {version of Penrose's retrodiction paradox.}

\section{Is there a Penrose retrodiction paradox in quantum mechanics in the first place?}\label{Theory}

\begin{quote}
``\textrm{[...]} \emph{it is reasonable to believe, in principle at least, that the theory would be adequate if only the calculational problems could be overcome}'' (Willis E. Lamb Jr \cite{Lamb}, page 23).
\end{quote}

The goal of this section is to show that an accurate application of the formalism of quantum mechanics actually leads to the resolution of  Penrose's retrodiction paradox in a simple and consistent manner. This will be achieved by calculating the \emph{fully} unitary (and, therefore, reversible) photon dynamics, from the emission to the detection, in  Penrose's exemplary system depicted in figure \ref{fig1} (a).

\subsection{Refinement of the model}\label{Model}

\begin{figure}[!ht]
%
\centerline{\includegraphics[scale=3,clip=false,width=.55\columnwidth,trim = 0 0 0 0]{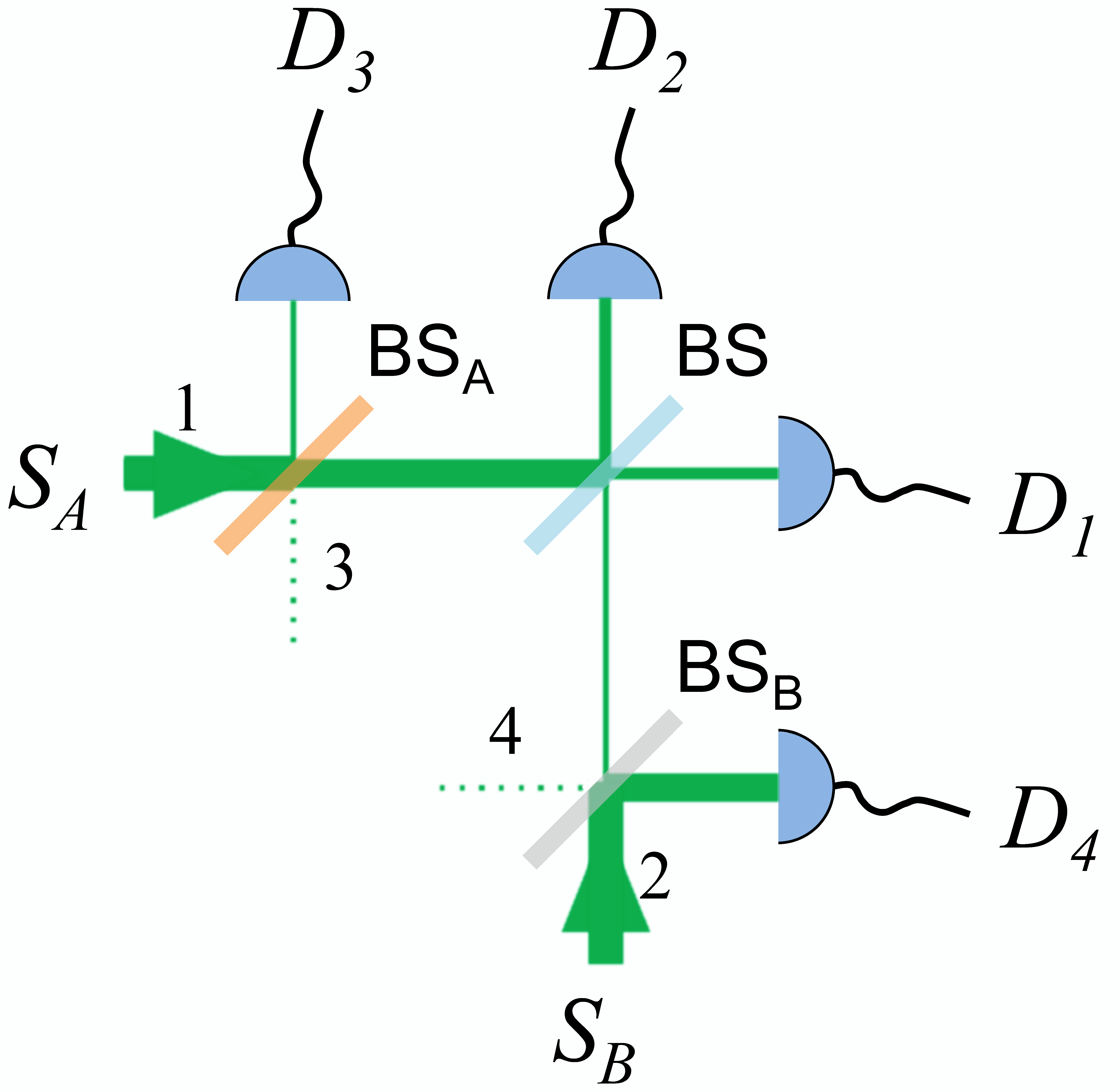}}
\caption{\label{fig3}
Amended version of Penrose's quantum retrodiction paradox.
Two deterministic  sources $S_A$ and $S_B$ emit single photons in modes $1$ and $2$ aimed at beam-splitters $\mathsf{BS_A}$ and $\mathsf{BS_B}$, respectively.  $\mathsf{BS_A}$ couples modes $1$ and $3$ with high transmittance $\abs{t_A}^2 \approx 1$ and low reflectance  $\abs{r_A}^2 = 1 - \abs{t_A}^2 $. $\mathsf{BS_B}$ mixes modes $2$ and $4$ with  high reflectance $\abs{r_B}^2 \approx 1$ and low transmittance  $\abs{t_B}^2 = 1 - \abs{r_B}^2 $.
Mode $1$ is identified with the straight line connecting the source $S_A$ with the detector $D_1$. Mode $2$ goes from $S_B$ to $D_2$. Mode $3$ connects the empty input port of  $\mathsf{BS_A}$ with $D_3$. Finally, mode $4$ joins the empty input port of  $\mathsf{BS_B}$ with $D_4$.
If transmitted (with high probability $\abs{t_A}^2$) through $\mathsf{BS_A}$,  photon $A$ enters the ${50\!:\!50}$ beam splitter $\mathsf{BS}$ via the input port $1$; if reflected (with low probability $\abs{r_A}^2$), it is counted by detector $D_3$. Conversely, if  photon $B$ is reflected (with high probability $\abs{r_B}^2$) by $\mathsf{BS_B}$, it reaches  detector $D_4$; if transmitted (with low probability $\abs{t_B}^2$), the photon enters $\mathsf{BS}$ from input port $2$. Source $S_A$ and beam splitter $\mathsf{BS_A}$ jointly  replace the random source $S$ in figure \ref{fig1} (a); source $S_B$ together with beam splitter $\mathsf{BS_B}$ are put in place of the floor $F$. The output ports $1$ and $2$ of $\mathsf{BS}$ are monitored by detectors $D_1$ and $D_2$, respectively. The former replaces detector $D$ in figure \ref{fig1} (a); the latter acts for the ceiling $C$. The varying thickness of the green lines represents, in a cartoon-like fashion, the transmittance and reflectance of the beam splitters. }
\end{figure}

To begin with,  let us consider again the experimental layout illustrated in figure \ref{fig1} (a). A short quantum-mechanical  description thereof  has already been provided in the introduction. However, as it will be clear soon, a \emph{thorough} quantum-mechanical description of this layout requires some amendments.  First of all, as Penrose contemplated the occurrence of the very improbable event of a photon emitted by the floor $F$, we have to account for such instance in a proper quantum-mechanical way. To this end, consider
 a deterministic single-photon source $S_B$  followed by a beam splitter $\mathsf{BS_B}$ with high reflectance $\abs{r_B}^2 \approx 1$ and low transmittance  $\abs{t_B}^2 = 1 - \abs{r_B}^2 \ll 1$, as shown in the bottom-right part of figure \ref{fig3} \cite{Campos}.
\\
When $S_B$ emits a photon, there is only a tiny probability $\abs{t_B}^2$ of the photon reaching the  beam splitter $\mathsf{BS}$. This process is thus equivalent to a photon being emitted by the floor with  probability $\abs{t_B}^2 \ll 1$.
To check whether the photon  has been actually transmitted towards $\mathsf{BS}$ or not, we monitor the output port $4$ of  $\mathsf{BS_B}$ with the detector $D_4$. When the last does not click, a photon is aiming at $\mathsf{BS}$.

The main source present in the setup of figure \ref{fig1} (a) is $S$. A realistic quantum-mechanical description of it must account for a conceivable limited emission efficiency and/or losses.  Again, we model such imperfect source $S$ as an ideal source $S_A$ followed by a beam-splitter $\mathsf{BS_A}$ with low reflectance $\abs{r_A}^2 \ll 1$ and high transmittance  $\abs{t_A}^2 = 1 - \abs{r_A}^2 \approx 1$, as shown in the top-left part of figure \ref{fig3}. When $S_A$ emits a photon, there is almost $100\%$ probability  of the photon reaching the beam splitter $\mathsf{BS}$ because $\abs{t_A}^2 \approx 1$.
Then, to test whether the photon  has been transmitted through  $\mathsf{BS_A}$ or not, we observe the output port $3$ of  $\mathsf{BS_A}$ with the detector $D_3$.

The last modification of the setup of  figure \ref{fig1} (a) is about the absorbing ceiling $C$. In order to achieve complete information about the state of the two photons emitted by $S_A$ and $S_B$, we must replace $C$ with the detector $D_2$ at the  output port $2$ of $\mathsf{BS}$; at a later stage we can always decide whether to use this additional information or not. At the other  port (number $1$) of the last, we put the detector $D_1$ which has the same role of $D$ in figure \ref{fig1} (a).

To summarize, a realistic quantum-mechanical representation  of the layout presented in figure \ref{fig1} (a)
 requires the presence  of two ideal deterministic sources $S_A$ and $S_B$, four field modes labeled $1$ through  $4$ and four detectors  $D_1$, $D_2$, $D_3$ and $D_4$  eventually coupled to the modes $1,2,3$ and $4$, respectively, as shown in figure \ref{fig3}. Sources $S_A$ and $S_B$ are assumed to emit single photons into modes $1$ and $2$. Mode $1$ is combined with mode $3$ by the beam splitter $\mathsf{BS_A}$ and mode $2$ is combined with mode $4$ by the beam splitter $\mathsf{BS_B}$. At last, modes $1$ and $2$ are coupled by the  beam splitter $\mathsf{BS}$ which, according to the example suggested by Penrose,  has reflection and transmission amplitudes  $r = i/\sqrt{2}$ and $t = 1/\sqrt{2}$, respectively. All beam splitters are supposed to be lossless
 \footnote{This is a further idealization that has not consequences upon our following reasoning because a lossy beam splitter can always be represented as a series of two consecutive ideal beam splitters. Then, provided that each unused port of the two beam splitters is monitored by a detector, complete information about the state of the two photons can always be obtained.}.

\subsection{Unitary photon dynamics}\label{Calcoli}

Let $m=1,2,3,4$ be an index labelling the four modes of the system shown in figure \ref{fig3}. Each mode can be populated by $n=0,1,2,\ldots,$ photons. When the mode $m$ is not excited ($n=0$) its quantum state is denoted as \emph{vacuum state} and is represented by the vector $\Psi_m^0$. When the mode is populated by either one or two photons its state is represented by $\Psi_m$ or $\Psi_m^2$, respectively. In the more abstract formalism of ``ket'' vectors and operators,  these state are described as
\begin{eqnarray}
\Psi_m^n := \frac{(\hat{a}^\dagger_m)^n}{\sqrt{n!}} \ket{0}, \qquad (n=0,1,2,\ldots),
\end{eqnarray}
where $\Psi_m^0:=\ket{0}$ \cite{LoudonBook}.
For the two sources emits single photons, modes $1$ and $2$ can be found excited in either the vacuum state $\Psi_m^0$ or the single- and the two-photon states $\Psi_m$ and $\Psi_m^2$, respectively, with $m=1,2$. However, because of the geometry of the setup, modes $3$ and $4$ can be  excited only in either the vacuum state $\Psi_m^0$ or the single-photon state $\Psi_m$, with $m=3,4$.

At generic time $t$ the electromagnetic field populating the four modes of our system can be described  by the state vector
\begin{eqnarray}
\Psi(t) =\Psi_1^{n_1} \otimes \Psi_2^{n_2}  \otimes \Psi_3^{n_3}  \otimes \Psi_4^{n_4},
\end{eqnarray}
where the symbol ``$\otimes$'' denotes the direct (or, Kronecker) product between vectors \cite{Weinberg} and $n_m \in \{0,1,2,\ldots\}$ gives the number of photons present in the mode $m$. This notation is precise but  somewhat cumbersome therefore, for the sake of clarity, hereafter we shall omit from our formulae both the symbol ``$\otimes$'' between the state vectors of different modes and the vector symbol  ``$\Psi_m^{n_m}$'' itself whenever $n_m =0$. Thus, for example, the state vector representing one photon in mode $1$ and one photon in mode $3$ will be concisely written as $\Psi_1 \Psi_3$ instead of $\Psi_1 \otimes \Psi_2^0  \otimes \Psi_3  \otimes \Psi_4^0$.

Now assume that the two sources $S_A$ and $S_B$ had emitted at $t=0$ one photon each in modes $1$ and $2$, respectively.
Using the notation introduced above,  we can write  the initial two-photon  state  $\Psi(0)$  as
\begin{eqnarray}
\Psi(0) =  \Psi_1 \Psi_2.
\end{eqnarray}
When the photons enters beam splitters $\mathsf{BS_A}$ and $\mathsf{BS_B}$ they either continue propagating along modes $1$ and $2$ or are reflected into modes $3$ and $4$. This is accounted for by the unitary evolution of the state vector
\begin{eqnarray}\label{Psi1}
\Psi(0) \rightarrow \Psi(1)& = & \; \left( \hU_\mathsf{BS_A}\Psi_1 \right) \left(\hU_\mathsf{BS_B}\Psi_2\right) \nonumber \\
&= & \; t_A t_B \Psi_1\Psi_2 + t_A r_B \Psi_1\Psi_4 + r_A t_B \Psi_2\Psi_3 + r_A r_B \Psi_3\Psi_4,
\end{eqnarray}
where $\hU_\mathsf{BS_A}$ and $\hU_\mathsf{BS_B}$  are  unitary operators (effectively, $2 \times 2$ matrices) representing the coupling between modes $1,3$ and $2,4$, respectively, with $\hU_\mathsf{BS_A}\Psi_1 =  t_A \Psi_1 + r_A \Psi_3$ and $\hU_\mathsf{BS_B}\Psi_2 = t_B \Psi_2 + r_B \Psi_4$ \cite{Jordan,Skaar}.

Next, it is not difficult to see that under the action of the ${50\!:\!50}$  beam splitter $\mathsf{BS}$ which couples modes $1$ and $2$ only, the vector state $\Psi(1)$ transforms into
\begin{eqnarray}\label{Psi2}
\Psi(1) \rightarrow \Psi(2)& = & \; \hU_\mathsf{BS} \Psi(1) \nonumber \\
&= & \; \frac{i}{\sqrt{2}} \, t_A t_B \left( \Psi_1^2 + \Psi_2^2 \right)
+  \frac{1}{\sqrt{2}} \, t_A r_B \left( \Psi_1\Psi_4 + i \Psi_2 \Psi_4 \right) \nonumber \\
& & \; + \frac{1}{\sqrt{2}} \, r_A t_B \left( i \Psi_1\Psi_3 +  \Psi_2 \Psi_3 \right) +  r_A r_B \Psi_3\Psi_4,
\end{eqnarray}
where the unitary operator $\hU_\mathsf{BS}$ acts upon $\Psi_1$ and $\Psi_2$ according to $\hU_\mathsf{BS} \Psi_1 \Psi_2  = i\left( \Psi_1^2 + \Psi_2^2 \right)/\sqrt{2}$, $\hU_\mathsf{BS}  \Psi_1 \Psi_2^0  =  \left( \Psi_1 + i \Psi_2 \right)/\sqrt{2}$  and $\hU_\mathsf{BS}  \Psi_1^0 \Psi_2  = \left(   \Psi_2 + i \Psi_1   \right)/\sqrt{2}$ \cite{Campos,LoudonBook}.
It is instructive to remark that  $\hU_\mathsf{BS} \Psi_1 \Psi_2$  describes either two photons in mode $1$ or two photons in mode $2$ because the two state vectors representing both photons transmitted and both photons reflected cancel each other. This is the well-known bunching, or coalescence, effect predicted by quantum mechanics when two indistinguishable photons enter a ${50\!:\!50}$  beam splitter \cite{Ou,Grangier,Toppel}.

\begin{figure}[!h]
%
\centerline{\includegraphics[scale=3,clip=false,width=.9\columnwidth,trim = 0 0 0 0]{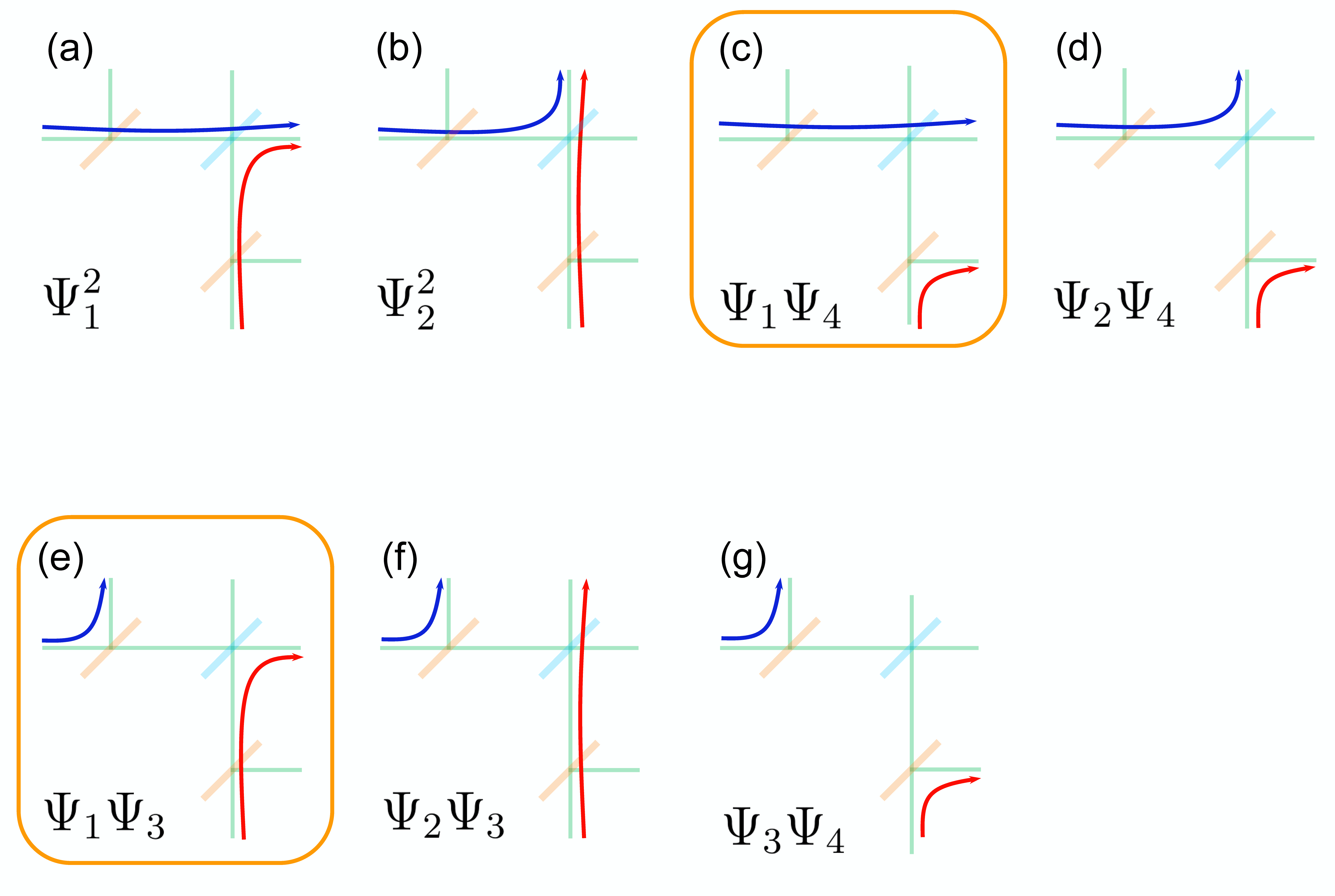}}
\caption{\label{fig4}
Diagrammatic illustrations of the mutually exclusive events described by the  seven terms in the expression \eref{Psi2} of $\Psi(2)$. Blue arrows highlight the paths followed by photons emitted from source $S_A$ in the setup of figure \ref{fig3}. Similarly, red arrows underline trajectories of photons emitted by $S_B$.
(a) Both photons from $S_A$ and $S_B$ reach detector $D_1$. This event happens with \emph{low} probability $\abs{t_A t_B}^2/2$. (b) The two photons emitted by   $S_A$ and $S_B$  arrive at detector $D_2$ again with  low probability  $\abs{t_A t_B}^2/2$. (c) The photons from $S_A$ and $S_B$ reach detectors $D_1$ and $D_4$, respectively. This occurs with \emph{high} probability   $\abs{t_A r_B}^2/2 \approx 50\%$. (d) Photon $A$ goes to $D_2$ and photon $B$ to $D_4$. The probability of this event is again $\abs{t_A r_B}^2/2 \approx 50\%$. (e) The photon from $S_A$ is reflected towards $D_3$ and the one from $S_B$ is reflected in the direction of $D_1$ with a low probability of  $\abs{r_A t_B}^2/2$. (f) The photons from $S_A$ and $S_B$ reach detectors $D_3$ and $D_2$, respectively. This event happens again with low probability $\abs{r_A t_B}^2/2$. (g) Photon $A$ is reflected towards $D_3$ and photon $B$ is reflected in the direction of $D_4$ with a low probability of $\abs{r_A r_B}^2$. The two framed panels (c) and (e) represent the only two events where one photon is detected by $D_1$.}
\end{figure}

The seven terms in the expression of $\Psi(2)$ are associated to specific and mutually exclusive physical processes, as illustrated in figure \ref{fig4}.
The first two terms of $\Psi(2)$  represent two photons directed at either $D_1$ or $D_2$ and, therefore, do not describe the instance considered by Penrose, namely \emph{one} photon aimed at $D_1$. The probability of occurrence of these processes is very small being $\abs{t_A t_B}^2/2 \approx \abs{t_B}^2/2 \ll 1$. The third and fourth terms  occur with high probability $\abs{t_A r_B}^2/2 \approx 1/2$ since $\abs{t_A}^2 \approx \abs{r_B}^2 \approx 1/2$, but only the third term proportional to $\Psi_1 \Psi_4$  describes a  photon directed at $D_1$, the case of our primary interest. The fifth and sixth terms describe processes taking place with \emph{very} low probability $\abs{r_A t_B}^2/2 \ll 1$ being both $\abs{r_A}^2 \ll 1$ and $\abs{t_B}^2 \ll 1$. It should be noticed that the fifth term precisely describes the process of ``a photon emitted by the floor and aimed at $D_1$'' that Penrose dubbed as ``absurd''. Differently from the elementary analysis presented in the introduction, here we can see that quantum mechanics does not forbid such a process but assigns to it a negligible, as opposed to the $50\%$ claimed in the introduction, probability of occurrence. Finally, the seventh term represents no photons aimed at either $D_1$ and $D_2$, an event that happens with low probability $\abs{r_A r_B}^2 \approx \abs{r_A}^2 \ll 1$.

\subsection{Measurement and discussion}\label{Detection}

In order to complete the quantum mechanical analysis of the retrodiction problem as formulated by Penrose, we must eventually consider the measurement process. Measurement theory plays a key role in the interpretation of quantum mechanics and,  after about a century from the foundation of the last,  is still a subject of fierce discussions amongst researchers. However, for the goal of the present work we will not need to join such debate and we simply address the interested reader to chapters 22 and 29 of \cite{Penrose} for a popular introduction to the matter, and to chapter 12 of \cite{PeresBook} and \cite{Gottfried} for a more technical exposition.

 Throughout this work we adopt the amended version of the celebrated von Neumann measurement theory \cite{Neumann} proposed by Moldauer \cite{Moldauer} and further elaborated by Peres \cite{Peres1,Peres2}. To make a long story short, according to this theory the measuring apparatus itself is considered as a quantum system and is characterized by a certain observable property represented by a given Hermitean operator.  Let  $\Phi_m^0, \Phi_m^1, \Phi_m^2, \ldots \,$, be the eigenstates of such operator characterizing the detector $D_m$.
Initially, each detector $D_m$ is prepared in the ground state $\Phi_m^0$ and each field mode $m$ is assumed to be in the state $\Psi_m^n$, representing $n$ photons in the mode $m$. Therefore, considering the photons and the detectors as a single composite quantum system, the initial state vector of the last is  given by the expression \eref{Psi2} of $\Psi(2)$ with each vector $\Psi_m^n$  replaced by $\Psi_m^n \otimes \Phi_m^0$.
The interaction of the detector with the photons is such that the detector remains in the state $\Phi_m^0$ if $n=0$ (no photons counted) and ends up in the state $\Phi_m^n$ if $n\neq0$ ($n$ photons counted). For our purposes we do not need to elaborate further on this point, however, what is crucial for our reasoning is that it is possible to show that such measuring process can be represented by a \emph{unitary operator} and is, therefore, perfectly reversible \cite{Moldauer,PeresBook}.
Thus, finally, we are ready for answering the question posed in the introduction:
\begin{quote}
\emph{Given that a photon has been detected by $D$ at some time $t=0$, what is the probability of an early emission event by $S$ at a sufficiently former time $t=T<0$?}
\end{quote}
There are different answers to this question depending on the amount of information at our disposal.
Specifically, here we consider the two cases where \emph{a}) we can read the counter of detector $D_1$ only; and \emph{b}) we have access to the counters of all detectors $D_1, D_2, D_3$ and $D_4$.
In the first case \emph{a}), the experimenter has at his disposal the minimal amount of information about the system.   Vice versa, in the second case  \emph{b}) the available information is maximal. Of course, one could also consider a third, intermediate  case  \emph{c})  where the accessible information is comprises between the minimal and the maximal. One explicit example will be given in the next subsection \ref{background}.
At this point it is useful to recall that by definition $\abs{t_A}^2 \approx \abs{r_B}^2 \approx 1$ and $\abs{r_A}^2 \approx \abs{t_B}^2 \approx 0$.
\begin{itemize}
  \item[\emph{a})]  Detector $D_1$ can count $0,1$ or $2$ photons per time.
  According to our previous analysis illustrated in figure \ref{fig4}, there are four alternative histories  \textsf{(b)}, \textsf{(d)}, \textsf{(f)} and \textsf{(g)} for $0$ photons detected at $D_1$.
 This process is dominated by  photon history \textsf{(d)} occurring when the photon emitted by the source $S_A$ is reflected towards the ceiling by the ${50\!:\!50}$ beam splitter $\mathsf{BS}$. The probability for this event is $\abs{t_A r_B}^2/2 \approx 50\%$. The probabilities for the remaining three histories \textsf{(b)}, \textsf{(f)} and \textsf{(g)} are  $\abs{t_A t_B}^2/2$,  $\abs{r_A t_B}^2/2 $ and  $\abs{r_A r_B}^2 $, respectively, all negligible.
Then, there are two alternative histories  \textsf{(c)} and \textsf{(e)} for the detection of a single photon at $D_1$. There is a probability $\abs{t_A r_B}^2/2 \approx 50\%$ for the photon history \textsf{(c)} and  a negligible probability $\abs{r_A t_B}^2/2 $ for the photon history \textsf{(e)}.
Finally, there is a small probability $\abs{t_A t_B}^2/2$ for the photon history \textsf{(a)} yielding to $2$ photons detected at $D_1$.
           Given our limited amount of information, this is all what we can say about our observations, namely we can only make a \emph{probabilistic} retrodiction. Such probabilistic character of quantum retrodiction was already remarked long time ago by Einstein, Tolman and Podolsky who wrote that ``[...] the principles of quantum  mechanics actually  involve  an  uncertainty  in  the   description  of  past   events which  is analogous to  the  uncertainty in  the prediction of  future events.''(p. 780 of \cite{ETP}).  However, and this is the main result of this work, contrary to Penrose's claim quantum mechanics \emph{does} give,  correctly,  different probabilities for the two alternatives \textsf{(c)} (photon emitted by the source) and \textsf{(e)} (photon emitted by the floor). Thus, if we register a single detection event at $D_1$, we can retrodict that the detected photon was emitted by either the source $S$ with probability
           $P[\mathrm{event} \, \mathsf{(c)}]=1/( 1 + \epsilon) \approx 1 - \epsilon$, or by the floor $F$ with (insignificant) probability  $P[\mathsf{\mathrm{event} \, (e)}]=\epsilon /( 1 + \epsilon ) \approx  \epsilon $, where we have defined $\epsilon \equiv \abs{r_A t_B}^2/\abs{t_A r_B}^2 \ll 1$.
  \item[\emph{b})] The seven events illustrated in figure \ref{fig4} are clearly mutually exclusive. This means that given a specific sequence $\{d_1,d_2,d_3,d_4 \}$ of counts delivered  by $D_1,D_2,D_3$ and $D_4$, respectively, we can \emph{deterministically} retrodict the photon history. For example, if the detectors deliver the string of counts $\{1,0,0,1 \}$, this uniquely selects photon history \textsf{(c)}.
\end{itemize}
From the analyses presented above for both cases of \emph{a}) probabilistic and \emph{b}) deterministic quantum retrodiction, we can thus confirm that:
\begin{quote}
\emph{There is no  Penrose's retrodiction paradox in orthodox quantum mechanics.}
\end{quote}
All in all, considering the results presented here and in subsection \ref{retroclassical}, we can conclude that a careful application of the fundamental principles of classical and quantum optics and a proper handling of the information extracted from measurements,  leads to a complete resolution of Penrose's retrodiction paradox.

\subsection{Some technical background}\label{background}

For the more mathematically-oriented reader, in the remaining part of this section we supply the rigorous quantum-mechanical calculation of the probabilities for the several photon histories considered in  case \emph{a}) above.
Given the  state vector $\Psi(2)$ we may define a \emph{density operator} $\hat{\rho}$ as
\begin{eqnarray}
\hat{\rho} =  \Bigl[\Psi(2) \Psi^\dagger(2) \Bigr],
\end{eqnarray}
where the  generic linear operator $\left[ \Omega \Omega^\dagger \right]$, called \emph{dyad}, is a  projector upon the state $\Omega$ \cite{Weinberg}. The density operator contains as much information about the system as the original state
$\Psi(2)$. However, in case \emph{a}) above we deliberately chose to ignore that part of the available information coming from the counts provided by detectors $D_2,D_3$ and $D_4$. Quantum mechanics deals with this loss of information by introducing the so-called \emph{reduced density matrix} $\hat{\rho}_1$ describing photons in mode $1$ solely, obtained from $\hat{\rho}$ by tracing with respect to the unobserved modes $2,3$ and $4$ \cite{PeresBook}:
\begin{eqnarray}
\hat{\rho} \rightarrow \hat{\rho}_1& = & \; \Tr_{2,3,4}\Bigl[\Psi(2) \Psi^\dagger(2) \Bigr]\nonumber \\
& = & \; \sum_{n=0}^2 P_n \, \Bigl[\Psi_1^n {\Psi_1^n}^\dagger \Bigr],
\end{eqnarray}
where $P_0, P_1$ and $P_2$ are the probabilities that detector $D_1$ had counted $0,1$ and $2$ photons, respectively.

A straightforward calculation gives
\begin{eqnarray}
P_0 & = & \; \frac{1}{2}\abs{t_A t_B}^2 + \frac{1}{2}\abs{t_A r_B}^2  + \frac{1}{2}\abs{r_A t_B}^2  + \abs{r_A r_B}^2, \nonumber \\
& = & \; \frac{1 + \abs{r_A r_B}^2}{2} ,
\end{eqnarray}
\begin{eqnarray}
P_1  =  \frac{1}{2}\abs{t_A r_B}^2 + \frac{1}{2}\abs{r_A t_B}^2,
\end{eqnarray}
and
\begin{eqnarray}
P_2 =\frac{1}{2}\abs{t_A t_B}^2 ,
\end{eqnarray}
with $P_0 + P_1 + P_2 =1$. From the beam splitters properties $\abs{t_A}^2 \approx \abs{r_B}^2 \approx 1$ and $\abs{r_A}^2 \approx \abs{t_B}^2 \approx 0$, it follows that $P_0 \approx P_1 \approx 1/2$ and  $P_2 \approx 0$. Therefore, we can write the reduced density matrix $\hat{\rho}_1$ as:
\begin{eqnarray}
\hat{\rho}_1 \approx  \frac{1}{2} \Bigl[\Psi_1^0 {\Psi_1^0}^\dagger \Bigr] + \frac{1 }{2}\Bigl[\Psi_1 {\Psi_1}^\dagger \Bigr],
\end{eqnarray}
which expresses the fact that detector $D_1$ counts either zero or one photon with $50\%$ probability.

With this formalism at our disposal, we are now ready to consider the third case  \emph{c}) mentioned in the previous subsection \ref{Detection}, occurring when we possess  a partial information about the system. Specifically, we consider the case when \emph{we know} that there is either one photon in mode $1$ after the beam splitter $\mathsf{BS_A}$, or one photon in mode $2$ after the beam splitter $\mathsf{BS_B}$. We acquire this information when either detector $D_4$ or detector $D_3$, respectively, had recorded one photon.  Then, according to figure \ref{fig4}, either histories \textsf{(c-d)} or histories  \textsf{(e-f)}, may occur. Having this additional information at our disposal, we must update the description of the system by replacing the \emph{pure} vector state $\Psi(1)$ given in equation (\ref{Psi1}),  with a suitably defined density operator describing the two mutually exclusive instances of having one photon either in mode $1$  or in mode $2$.  When detector $D_4$ clicks and $D_3$ does not, the system is described by the state vector $\Psi(1)$ projected upon $\Psi_3^0 \Psi_4$, which represents the occurrence of zero photons in mode $3$ and one photon in mode $4$. Analogously, when detector $D_3$ clicks and $D_4$ does not, we must project $\Psi(1)$ on $\Psi_3 \Psi_4^0$. Therefore, after that one photon has been detected by either $D_4$  or $D_3$, our system will be represented by either the vector state
\begin{eqnarray}
 \Psi(1)& \to & \; \bigl(\Psi_3^0 \Psi_4, \Psi(1) \bigr) = t_A r_B \Psi_1,
\end{eqnarray}
with high probability $\abs{t_A r_B}^2 \approx 1$, or by
\begin{eqnarray}
 \Psi(1)& \to & \; \bigl(\Psi_3 \Psi_4^0, \Psi(1) \bigr) = r_A t_B \Psi_2,
\end{eqnarray}
with low probability $\abs{r_A t_B}^2 \approx 0$. In the mathematical formalism of quantum mechanics, these two instances can be properly represented by the normalized density operator
\begin{eqnarray}
 \Psi(1)& \to & \; \hr(1) =  W_1  \Bigl[ \Psi_1 \Psi_1^\dagger \Bigr] + W_2 \Bigl[\Psi_2 \Psi_2^\dagger \Bigr],
\end{eqnarray}
where we have defined
\begin{equation}
W_1 =  \frac{\abs{t_A r_B}^2}{\abs{t_A r_B}^2 + \abs{r_A t_B}^2} \equiv \frac{1}{1+\epsilon} , \quad W_2 = \frac{\abs{r_A t_B}^2}{\abs{t_A r_B}^2 + \abs{r_A t_B}^2}  \equiv \frac{\epsilon}{1+\epsilon},
\end{equation}
with $W_2 = 1- W_1$ and $\epsilon \equiv \abs{r_A t_B}^2/\abs{t_A r_B}^2 \ll 1$. In practice, the operator $\hr(1)$ tells us that when only one photon is present in our system, then it has been either emitted  by the source $S$ with probability $W_1 \approx 1 - \epsilon$, or by the floor $F$ with probability $W_2 \approx \epsilon$. It is important to remark that passing from the \emph{complete} description furnished by $\Psi(1)$, to the \emph{incomplete} one given by $\hr(1)$,  we could actually reduce our two-photon version of Penrose's paradox to the one-photon version originally considered by Penrose, which then manifests as a particular case of a more general problem.
However, also for this case our treatment shows that quantum mechanics furnishes the correct values for the retrodiction probabilities.

This calculation concludes our quantum-mechanical analysis of the retrodiction paradox as presented by Penrose.

\section{Summary and conclusions}\label{Conclusions}

\begin{quote}
``\emph{Quite simple, my dear Watson}'' (Sherlock Holmes,  \textsc{The adventure of the Retired Colourman}, Sir Arthur Conan Doyle).
\end{quote}

In this work we have proposed a simple solution of  {a specific interpretation of}
 the so-called  retrodiction paradox which manifests in both contexts of classical and quantum optics. Specifically, we have focused on Penrose's version of this phenomenon as given in his celebrated book
\emph{The Road to Reality} \cite{Penrose}.
A thorough analysis of this problem  eventually led us to the conclusion that there is no retrodiction paradox in optics, at least not in the form suggested by Penrose. In fact, we demonstrated that the deceptive inconsistencies arising in the quantum and classical descriptions of light when making retrodictions, are simply due to an incomplete knowledge of the state of the light. Why incomplete? Because light, either classical or quantum, in many instances does not behave as just being   ``here \emph{or} there'', but rather as being  ``here \emph{and} there''. This non-locality  is peculiar of all wave phenomena, irrespective of their  classical or quantum nature.  Therefore, when we measure some observable property of light only ``here'' (as it is the case of Penrose's paradox), we are missing all that part of information carried by the light ``there''. A typical example thereof is precisely given by the light entering beam-splitter \textsf{BS} in figure \ref{fig1} (a). When leaving the beam splitter, the light is both transmitted towards the detector $D$ \emph{and} reflected in the direction of the ceiling $C$, and this is true even at single-photon level\footnote{There is one caveat: While for bright light this sentence is perfectly correct, some caution must be applied when dealing with single photons. If with the word ``photon'' we actually denotes a ``detected photon'', then the latter is either here \emph{or} there. However, if instead with ``photon'' we designate a quantum of the electromagnetic field, then the latter may have a nonzero mean value  here \emph{and} there.}.
 In summary, whenever we possess sufficient information about our system, the use of either classical or quantum descriptions of light always leads to correct retrodictions  in the scheme proposed by Penrose.

In conclusion, paraphrasing a famous Danish proverb often quoted by Niels Bohr, we would like to state that \emph{retrodiction is not very difficult, especially about the past}.

\section*{Acknowledgements}

We are indebted to Robert B. Griffiths for insightful comments and useful suggestions.
AA thanks Lev Plimak for useful discussions and for having provided reference \cite{Lamb}

\end{document}